# DIANA Scheduling Hierarchies for Optimizing Bulk Job Scheduling

Ashiq Anjum[1, 3], Richard McClatchey[1,] Heinz Stockinger[2], Arshad Ali[3],
Ian Willers[4], Michael Thomas[5], Muhammad Sagheer [3], Khawar Hasham[3], Omer Alvi[3]

[1] CCS Research Centre, University of the West of England, Bristol, UK
[2] Swiss Institute of Bioinformatics, Lausanne, Switzerland
[3] National University of Sciences and Technology, Rawalpindi, Pakistan
[4] CERN, European Organization for Nuclear Research, Geneva, Switzerland
[5] California Institute of Technology, Pasadena, California, USA

## Abstract

*The use of meta-schedulers for resource management in large-scale distributed systems often leads to a hierarchy of schedulers. In this paper, we discuss why existing meta-scheduling hierarchies are sometimes not sufficient for Grid systems due to their inability to re-organise jobs already scheduled locally. Such a job re-organisation is required to adapt to evolving loads which are common in heavily used Grid infrastructures. We propose a peer-to-peer scheduling model and evaluate it using case studies and mathematical modelling. We detail the DIANA (Data Intensive and Network Aware) scheduling algorithm and its queue management system for coping with the load distribution and for supporting bulk job scheduling. We demonstrate that such a system is beneficial for dynamic, distributed and self-organizing resource management and can assist in optimizing load or job distribution in complex Grid infrastructures.*

## 1. Introduction

The Grid concept was created to facilitate the use of available distributed resources effectively and efficiently. The first step needed before one can utilize the Grid for running jobs is to locate and use (the best) resources available to serve those jobs i.e. resource scheduling. Applying the concept of P2P systems to resource scheduling, can lead to efficient resource utilization. Existing scheduling systems e.g. [1][2], are often based on the client-server architecture with one or several meta-schedulers [3][4] on top of independent local schedulers such as LSF, PBS etc. Each local scheduler can collect information and can schedule the jobs within its own managed site. Typically, these local schedulers cannot schedule jobs to some other available site.

Peer-to-Peer (P2P) scheduling systems on the other hand can provide environments where each peer can communicate with all other peers to make "global" decisions, can propagate their information to other peers, and can control their behaviour through this information. This feature should make scheduling decisions more efficient. In contrast to this P2P approach, centralized scheduler management can be problematic in several ways since load balancing, queue management, job allocation, policies etc. are central and are typically managed by a (single) central meta-scheduler and might not be fault tolerant. Note that by client server architecture*, we do not mean here a tier system* which uses various tiers, which are clients of each other, to scale up the client server behaviour. Each tier is not scaleable if treated in isolation.

Our intention is to incorporate a P2P approach so that schedulers do not take global decisions at a single central point, but rather many sites participate in the scheduling decisions. Each site should have information on load, queue size etc., should monitor its processing nodes and then propagate this information to other peers. *Local* and certain *global* policies could be managed at the site level instead of a central hierarchical management. As a result, the P2P behaviour can become an important architectural model for fault tolerant, self-discoverable and autonomous global resource scheduling.

Schedulers may be subject to failure or may not perform efficient scheduling when they are exposed to millions of jobs having different quality of service needs and different scheduling requirements. They may not be able to re-organize or export scheduled jobs which could result in large job queues and long execution delays. For example in High Energy Physics (HEP) analysis a user may submit a large number of jobs simultaneously (this being referred to as bulk job scheduling), and the scheduling requirements of bulk jobs may well be different to those of singly queued jobs. In bulk job submission by a single or multiple users at a particular site it might become impossible for a local scheduler to serve all the jobs without using some job export mechanism. In the absence of this mechanism, it is possible that some of the jobs might be lost by the scheduler. What is required is a decentralized scheduling system which not only automatically exports jobs to its peers under potentially severe load conditions (such as with bulk jobs), but at the same time it manages its own scheduling policies, whilst queuing jobs and monitoring network conditions such as bandwidth, throughput and latency. The queuing mechanism that is needed at each scheduling peer should follow a recognised management scheme. It should associate priorities to each job inside the queue, depending on the user profile and job requirements with the scheduler servicing high priority jobs preferentially to optimise Grid service standards. In this paper, we explain the functionality of a **P2P meta-scheduler** and present its scheduling and queue management mechanism and demonstrate the advantages and drawbacks of such a system implementation.

## 2. Hierarchy of Schedulers

A *meta-scheduler* coordinates communication between multiple heterogeneous *local schedulers* that typically manage clusters in a LAN environment (cf. Figure 1). In addition to providing a common entry point, a meta-scheduler also enables global access and coordination, whilst maintaining local control and ownership of resources through the local schedulers. The fundamental difference between a meta-scheduler and local schedulers is that a meta-scheduler does not own the resources and has no autonomy in its decisions. Therefore, the meta-scheduler does not have total control over the resources. Furthermore, a meta-scheduler does not have control over the set of jobs already scheduled to a local scheduler (also referred to as local resource management system). This lack of ownership and control are the sources of many of the problems to be solved in the meta-scheduling domain.

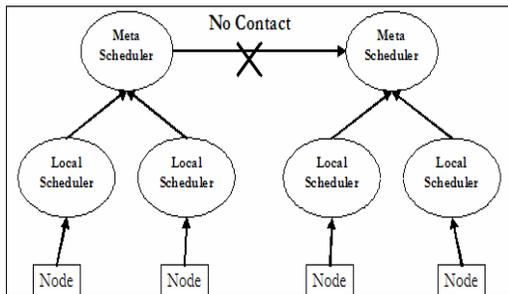
Fig. 1: No communication between meta-schedulers

In this example local and meta-schedulers form a hierarchy and individual schedulers sit at different levels in the hierarchy. Each local scheduler can cooperate and communicate with its siblings through a meta-scheduler, however each meta-scheduler cannot communicate with other meta-schedulers of other sites or Grids as shown in Figure 1. Communication is only possible between local schedulers and the meta-scheduler.

A user submits a job to a meta-scheduler (local to the user, typically at the same *site*) which in turn contacts a local scheduler. A particular meta-scheduler considers only its own managed sites to schedule the job and does not look around for other sites managed by other schedulers to distribute load and to get the best available resources. The jobs are scheduled centrally irrespective of the fact that this may lead to a poor quality of service due to long queuing and scheduling delays. Hence, the architecture with non-communicating meta-schedulers can lead to inefficient usage of Grid resources. Further, in this architecture the meta-scheduler schedules the job on its site, cannot communicate with the sibling meta-schedulers and hence does not consider the underlying network and data transfer costs between the sites. This is one of the reasons that almost all Grid deployments have at most only a few meta-schedulers and that any two cannot communicate and interoperate with each other. In contrast, in this paper we discuss the DIANA scheduling algorithm and how it is achieved through a P2P meta-scheduling hierarchy, and describe the underlying mathematical and implementation details for managing queue and load balancing in the DIANA meta-scheduler.

This approach is not simply an 'all-to-all' communication. The nodes are managed by local schedulers which report to the site meta-schedulers. The site to site communication is in essence a P2P communication between meta-schedulers. Each meta-scheduler maintains a table of entries about the status of the local schedulers, the queue length, jobs in execution mode, and the nodes managed by them which is updated in real time when a node joins or leaves the system. When a user submits a job, the site meta-scheduler communicates within the local scheduler to find the suitable resources. If the required resources are not available within the site, it contacts the meta-schedulers of other sites in the virtual organisation (VO) which have suitable resources. This approach is thus not just all-to-all communication and involves a reduced set of message passing between the meta-schedulers. Furthermore, communication between the meta-schedulers is not very frequent, meta-schedulers communicate only after fixed intervals to update the status of their resources to each other. A meta-scheduler might also require to communicate if a group of jobs at a site needs to be exported to a site having better resources. Therefore, this meta-scheduler communicates with other meta-schedulers for load evaluation and cost determination for job submission to that remote site.

## 3. Meta-scheduling with DIANA

It is important in Grid systems to have a distributed meta-scheduler, which implements the features discussed in Section 2, and that site meta-scheduler instances should interoperate and communicate with each other, should be fault tolerant and self organizing and should make network aware (which includes network characteristics in the scheduling decisions) data intensive decisions. In addition to being network-aware, the meta-scheduler should avoid making centralized decisions. It should communicate and share the information with all other meta-schedulers so that Grid resources are well evaluated and utilized.

DIANA is a Data Intensive and Network Aware meta-scheduler which performs global meta-scheduling in a local environment, typically in a LAN. In DIANA, we do not use independent meta-schedulers but use a set of meta-schedulers that work in a P2P manner. Each site has a meta-scheduler that can communicate with all other meta-schedulers on other sites as shown in Figure 2. The scheduler is able to discover other schedulers with the help of a P2P discovery mechanism [5]. We do not replace the local schedulers in this architecture, rather we have added a layer over each local scheduler so that site meta-schedulers can talk directly to each other instead of getting directions from a central global meta-scheduler.

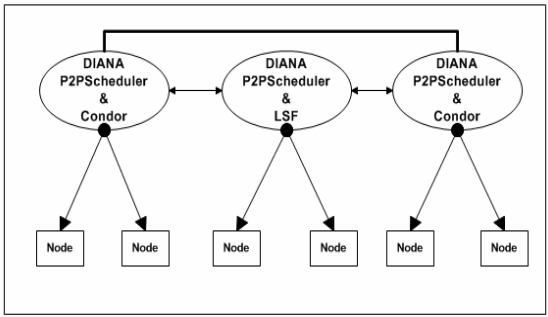

Fig. 2: P2P Communication between Schedulers

A meta-scheduler can thereby obtain information from any other site and can make global decisions. Local information includes processing power, memory, site load, and queue length and network capability. The meta-scheduler will make scheduling decisions based on three essential factors: the network cost, the computation cost and the data transfer cost [6]. It can communicate with other meta-schedulers and may transfer jobs to other sites. It may transfer a job along with its required data to a remote site, consequently it should also consider the estimated transfer time of the job and data to that particular remote peer. Before making the scheduling decision, it should also consider the estimated computing capabilities of remote peers. Hence, the job will be submitted to the site with the least total cost.

In DIANA, the P2P behaviour is complemented by a discovery service. This discovery service maintains a list of available/alive peers in different ways. One way is that whenever a peer meta-scheduler is introduced to the network, it will inform the discovery service about its availability and when a peer is properly shutdown, it will update the discovery service about its new status. This leads to the question: what would happen if a peer suddenly went down without informing the discovery service? In order to cope with this issue, the discovery service uses an echo request/reply communication with the peers currently available in the list. The peer which does not reply is simply removed from the list. Each meta-scheduler site periodically contacts a discovery service to collect the updated information about the available peers. After getting this information, the peers start communicating with other meta-scheduling peers and update their local repositories with this information.

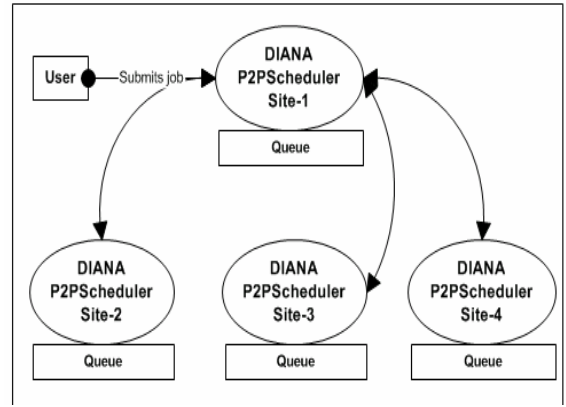

Fig. 3: Queue and DIANA Meta-Scheduler Instances

## 4. Queue Management

In conventional client-server scheduling architectures, local schedulers handle their queues at the site level whereas a meta-scheduler has a global queue at some central location. However, in the DIANA architecture, there is one DIANA meta-scheduler at each site, i.e. the DIANA P2P meta-scheduler layer sits on the top of one or many local schedulers at each site. In client-server architecture such as the one used by the gLite meta-scheduler, there is only one large queue at the meta-scheduler with local queues at each site. However, in the proposed P2P architecture each site meta-scheduler has knowledge about the local queue (s) plus a global queue which is managed by the DIANA layer. This leads to a scalable and self-organizing meta-scheduling behaviour which was missing in some of the conventional client-server scheduling architectures.

Each meta-scheduler has a queue management mechanism where it can queue the incoming jobs in a *Scheduler Queue* as shown in Figure 3, and the meta-scheduler assigns priorities to the incoming jobs. In Grid scheduling we have "user quotas" (user quota is the number of jobs a user can submit within a definite period of time), network characteristics, data locations and securely granted user privileges and therefore, each meta-scheduler needs to maintain its queue according to these criteria.

Before migrating a job, questions need to be answered such as:

- "What is the queue length on the target site?"
- "Can the target site execute the job quicker than the current site?"
- "If the job is migrated to another site, what will be the job priority on the remote site?"
- "How many jobs are ahead of this job in terms of priority?"

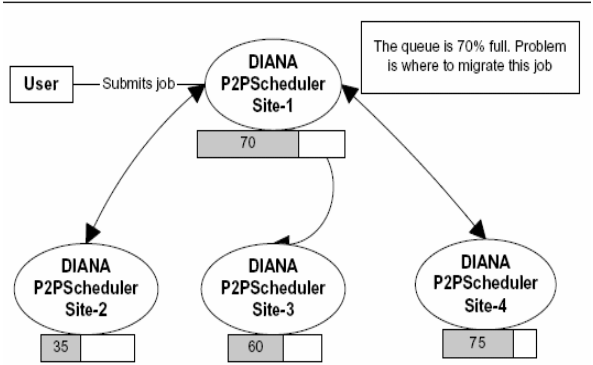

Fig. 4: Queue Management in DIANA P2P Meta-Scheduler

These considerations can have a significant effect on Grid performance. Figure 4 illustrates this queue management issue in the DIANA meta-scheduler. At each site there are two queues. One is the DIANA queue and the other is the queue of the local resource management system. *Only* the jobs from the DIANA meta-scheduler queue are exported to other sites. In contrast, once a job is *allocated* to a *local scheduler* at a site, it is never exported and waits in the local queue until assigned to a processor. All the prioritization of jobs, policy enforcement, migration and job steering issues are handled at the DIANA P2P level whereas the local scheduler works exactly in the same fashion as before once the job has been allocated to it.

## 5. Scheduling Algorithm

There are two scheduling schemes that the proposed algorithm will use: a **Normal Scheduling Scheme** and **Job Migration** (see Section 6). The Normal Scheduling Scheme is applicable to those jobs which have arrived for scheduling for the first time and have not, as yet, been migrated. Here, the meta-scheduler consults its peers, collects information about the peers (including network, computation and data transfer) and selects the site having minimum cost. It selects whichever site is the best site for its execution based on this cost estimation scheme. The meta-scheduler deals with both computational jobs and data intensive jobs using the DIANA meta-scheduling algorithm:

1. In the case of *computational jobs* (i.e. the job requires mainly CPU time), the meta-scheduler should schedule a job to the site where the computational cost is a minimum. At the same time, we have to transfer the job's files so that the job can be transferred as quickly as possible. The job might also require some input data which suggests selecting a site which has better network capacity (i.e. highest response time and lowest latency). Therefore, the meta-scheduler will select the site with minimum computational cost and but also takes into account the data transfer cost.
2. In the case of *data intensive jobs*, our preferences will change. In this case our job has more data and fewer computation requirements, and we need to identify the site where data can be transferred quickly and where computational cost is also low. In this case, data location will play an important role since data the transfer cost will be the key element in such a scheduling decision.
3. In most cases jobs are at the same time both compute as well as data intensive and will most likely follow the third category of the algorithm. In the third category the algorithm considers compute cost, network cost, data location and data transfer costs, and the site having *minimum aggregate cost* is selected for job execution.

## 6. Job Migration Algorithm

Consider a scenario in which a user submits a job to a meta-scheduler which places the job in a queue. If the queue management algorithm (see Section 7) of the meta-scheduler decides that this job should remain in the queue, it may have to wait some time before it gets scheduled or before migration to another site. The *Queue Management Module* of the meta-scheduler will ask the *Scheduling Module* to migrate this job. One important point to be noted here is that we want to locate the site where this job can be executed earliest. Consequently, our peer selection criterion is based on two things: a **minimum queue length** and **a minimum cost** to execute this job on the remote site. The meta-scheduler will communicate with its peers and will ask about their current queue length and the number of jobs ahead of this job. The site with the minimum queue length and minimum total cost is considered as the best site to where the job can be migrated.

Firstly, the algorithm will get the information about the available peers from the discovery service. Then it will communicate with each peer and collect the peer's queue length, the total cost and number of jobs ahead in terms of job priority. Then it will determine the site with the minimum queue length and the minimum jobs ahead. If the number of jobs and the total cost of the remote site is higher than the local cost, then this job is scheduled to the local site (i.e. it will *not* be migrated). If other sites are congested then there is no benefit in migrating the job, and that job will remain in the local queue and will eventually be served on the local site. Otherwise, the job is moved to the remote site, subject to a cost mechanism. Note that the DIANA meta-scheduler does not consider each job for export process rather a group of jobs is exported to a remote site which can significantly save the execution time on the remote site. It will not be cost effective to poll the remote peers and collect the queue and cost information for each job. This process is only carried out for bulk jobs or groups of jobs which are likely to take more time on their local sites. Furthermore, this will reduce communication traffic between the peers since all peers are polled only after some intervals when jobs at the sites need to be exported. Otherwise if all peers are polled for each job, this would significantly increase the communication traffic between peers.

## 7. Queue Management for Bulk Scheduling

In DIANA we propose a **multi-queue, feedback-oriented queue management** approach for **bulk job scheduling**. Users may send jobs in a burst, and the meta-scheduler has to place all these jobs in queues after assigning priorities. We must ensure that the priority of the jobs *decreases* as the number of jobs in the queues from a particular user *increases*. This is important otherwise a single user may send thousands of jobs in a burst and thereby improve the priorities for all his jobs.

In a typical scenario, when a user submits a job at a site for execution, the job is first placed in one of the queues managed by the queue management module of the meta-scheduler. A reprioritization algorithm may result in the migration of jobs from low priority to high priority queues or from high priority to low priority queues. The reprioritization technique eliminates the need for aging since the jobs are assigned new priorities on the arrival of each new job, and each job gets its place in the queues according to its new circumstances.

In the case of congestion at the site, the queue management algorithm will migrate the jobs to any other remote site where there are fewer jobs waiting in the queues. Note that only low priority jobs are migrated to remote sites since low priority jobs will have to wait for a long time in the case of congestion. Knowing the arrival rate (job submission rate) and service rate (job execution rate) of the jobs, we can decide whether or not to migrate the job to some other site. The formula [7] to decide whether there is congestion in the queues or not is simply:

*If ((Arrival Rate – Service Rate ) / Arrival Rate) > Thrs*

where Thrs is the threshold value configurable by the administrator. If we increase Thrs, then this means that the arrival rate exceeds the service rate and we allow more jobs in the queues. In any case this value will lie in the [0, 1] interval. Taking this, we can now explain the queue management algorithm. The job's place in the queue will be determined by the priority associated with the job, which is calculated by taking into account the quota of the user submitting the job, the execution time required by the job and the threshold on the number of jobs by a user.

Each queue will contain jobs having priorities falling in its specified priority range. According to our priority calculation algorithm, the priority of all the jobs will be in the interval [-1, 1] where -1 indicates the lowest priority and 1 indicates the highest priority. In the process of selecting the job's position in the queue, we place the jobs in descending order of their priorities i.e. the highest priority job will be placed first in the queue and the same order is followed for the rest of the jobs.

Suppose 'n' is the total number of jobs of the user in all queues, including his new job. Let the new job require 't' processors for the computation and 'T' be the total processors (including 't') required by all the jobs present in all queues. We denote the quota of the user, submitting the new job, by 'q' and the sum of the quotas of all the users, currently having their jobs in the queues including 'q', by 'Q'. Therefore, if the new user has already some jobs in the queues, 'q' will appear just once in 'Q'. Let 'L' be the sum of lengths of all queues, i.e. the total number of jobs present in all queues including the new job. Thus, if there are already 1500 jobs in the queues when 100 new jobs arrive, L = 1600. To assign a new job a place in the queue, we associate a number to it. This number is called the "priority" of the job and has its value in the interval [-1, 1]. The rule is that "the higher the priority, the better placed the job will be". Obviously, if priority is in the range [0,1], it will be considered good. To attain a good priority we must meet the following two constraints:

$$\frac{n}{L} \leq \frac{q}{Q} \quad \text{and} \quad \frac{1}{L} \geq \frac{t}{T} \quad \ldots\ldots (1) \text{ or}$$

$$n \leq \frac{(q \times L)}{Q} \quad \text{and} \quad L \leq \frac{T}{t} \quad \ldots\ldots (2)$$

Combining these inequalities (1) and (2), we get

$$n \leq \frac{(q \times T)}{(Q \times t)} \quad \ldots\ldots (3)$$

We denote $\frac{(q \times T)}{(Q \times t)}$ by 'N where 'N' represents the threshold and is clearly dynamic. For each job, its value will be different. If a user's number of jobs in the queue crosses this threshold then the priority of the jobs crossing the threshold 'N' must be lowered. The following algorithm calculates the priority of the new job:

*If( n <= N )*
    *Pr(n) = (N – n) / N*
*Else*
    *Pr(n) = (N – n) / n*

where Pr(n) denotes the priority of the new job. On the arrival of each job, the priorities of all the other jobs will be calculated again. This technique is known as *reprioritization*. The reason for doing this that we want to make sure that the jobs encounter minimum average wait time and the most 'deserving' job in terms of quota and time is given the highest priority. Moreover, by using this strategy we do not need to worry about the starvation problem and there is no need for aging since jobs are reorganized on the arrival of each new job.

## 8. Results and Discussion

We present here a performance comparison conducted using the DIANA P2P meta-scheduler which is a Web service and uses a Grid services framework called *JClarens* [8] to deploy this service.

We implemented a classical scheduling algorithm which works in a round-robin manner to compare it with the DIANA P2P meta-scheduler for job scheduling on various sites. Henceforth, we will refer to it as a 'Round Robin Scheduler' or 'Simple Scheduler'.

For simplicity we have used our own test Grid (rather than a production environment) to obtain results since a production environment requires the installation of many other Grid components that are not required for our experiments. We used five sites located in Pakistan (NUST), Switzerland (CERN) and the UK (UWE) for the purposes of our tests. Site 1 has four nodes, and the remaining four sites have five nodes each.

Two types of jobs are used in these experiments. One type of job is *compute intensive* which is a simple prime number calculator (between a specified range) and the other is *data intensive physics analysis* job which requires large amounts of data as input but performs computation over this data as well.

### 8.1 DIANA With Single Queue

In our first experiment, we submitted 1000 compute intensive jobs and calculated their execution times. Condor is used as a local scheduler for all of our tests. The execution time includes the time required to schedule and execute the job to one of the 'best sites' plus the time required in sending the data and job to that remote site. The scheduling decision made by DIANA in this experiment is independent of the queue mechanism (Shortest Job First (SJF) or Priority based queue) and therefore the first experiment uses a single queue.

As shown in Figure 5, the DIANA P2P meta-scheduler has a significantly better execution time compared to the 'Round Robin Scheduler' algorithm. We then calculated the queue times of the jobs (cf. Figure 6) to compare how effectively DIANA can reduce the wait time. The queue time here is the sum of the time in the meta-scheduler queue and the time spent in the queue of the local resource manager. Sometimes the queue time is even greater than the execution time if the resources are scarce compared to the job frequency. In this case compute intensive bulk jobs are placed in the queue before the DIANA meta-scheduler allocates them to the appropriate sites. The queue is maintained on a FCFS basis.

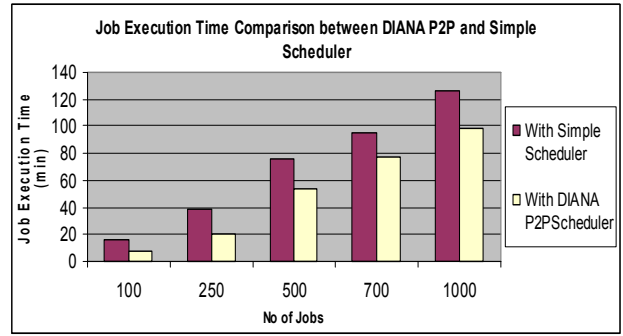

Fig. 5: Execution Times

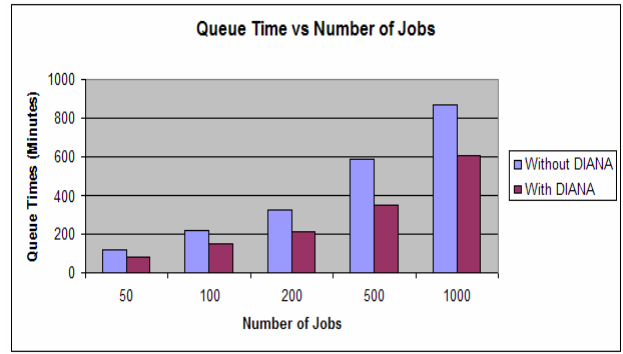

Fig. 6: Queue Times

### 8.2 DIANA With Multiple Queues

After this we submitted a set of jobs to calculate the job execution time with a SJF Queue Mechanism. These compute intensive jobs are similar with respect to their prime number requirements but they are different with respect to their inputs since each job has a different input range. These jobs are of varying processor requirements such as 8, 17, 26, and 35. The job demanding 8 processors has an input range 1-19999, 17 processors job has an input range 1-99999, 26 processors job has an input range 1-444444, and 35 processors job has input range 1-555555.

All jobs are submitted to the scheduler, which arranges them in its queue in an SJF manner on the basis of the job's processors requirement. In the comparison graph of Figure 7, it is clear that the performance (i.e. the execution time of jobs) gained via DIANA is much better than that of the 'Round Robin Scheduler' algorithm. The reason behind this is that DIANA worked on a SJF basis which reduces the total execution time since short jobs do not have to wait for long jobs. Similarly, the DIANA P2P meta-scheduler with its multi-queue priority mechanism has a better execution time compared to the 'Simple Scheduler' algorithm. Multi-queues not only enable the 'short job first' execution but also manage the queues on a priority basis, and this mechanism has significantly reduced the total execution times. From the data collected in these experiments, we can easily decide which approach is best in terms of job execution time. From Figure 7 it is clear that our priority driven approach results in more efficient execution time than other approaches.

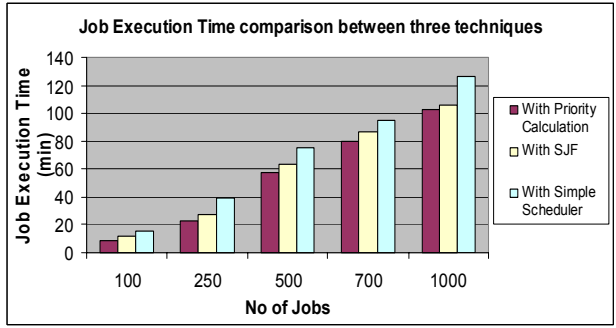

Fig. 7: Execution time Comparison

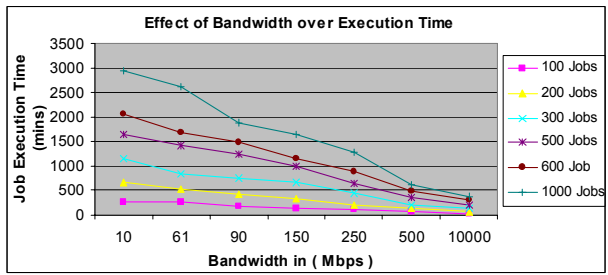

Fig. 8: Execution times vs. Bandwidth

### 8.3 Data Intensive Jobs and Network Issues

The rest of the results are related to network issues and have more impact on data intensive jobs than compute intensive jobs. In this experiment, we submitted the same number of jobs to different sites with different network conditions. The bandwidth varied from 10 Mbps to 1000 Mbps so that we can gauge its effect on the job execution time. We used Iperf to generate the extra network traffic and saturate the network so that available bandwidth can vary from 10 to 1000 Mbps. In these tests we show the effect of bandwidth on the execution time of the job. The data size is 10 GB and is the same for all jobs.

We see that bandwidth plays a vital role in scheduling decisions. The 'Round Robin Scheduler' algorithm will schedule the job to one of the sites without consulting the network conditions of that site. This approach will cause the user additional wait time since more time is consumed in transferring the executable and the data. In our proposed approach, before making any scheduling decisions for data intensive jobs, the network and bandwidth parameters are considered to select the best sites and we can see the impact of this approach in Figure 8. A lower bandwidth can often result in higher network costs, and the increase in network cost also affects the overall performance of the distributed system. Figure 8 is the comparison of different network costs against the execution time. Bandwidth is the only significant parameter in the network cost therefore we draw the execution time against the bandwidth.

### 8.4 Scalability Tests

In conclusion we present here the results of the scalability tests for the DIANA scheduling approach. These are *simulation* results since it was not feasible to deploy the DIANA system on such a high number of sites. In these tests, we assumed that there is a meta-scheduler on each node (here, a node corresponds to a site), and all the nodes work in a P2P way.

As shown in Figure 9, the number of nodes/sites and the number of jobs scheduled to the Grid was increased gradually to test which algorithm gives the steepest increase in time taken. An exponential increase is "bad" behaviour and shows that the algorithm is not scalable. In this test, jobs of a processing requirement of 3 MFLOP and a bandwidth load of 1 MB are launched to the Grid.

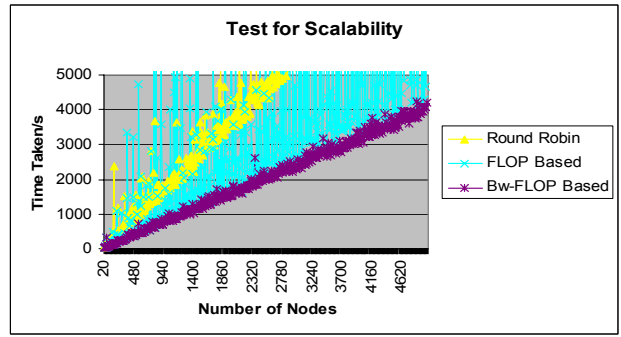

Fig.9: Scalability of the DIANA approach

The 'Round Robin Scheduler' algorithm has a steep linear curve showing that it is the most unscalable of the candidates. FLOP ((Floating Point Operations per Second) is a common measurement for the computational capability of a computer. A FLOP based algorithm could be considered as being completely opposite to the 'Round Robin Scheduler' algorithm, since it tries to gain complete knowledge about the current state of resources so that it can schedule jobs to the most powerful available machine, guaranteeing the quickest possible runtime. FLOP shows far too much variation in this case, although it clearly is more scalable than round robin. The DIANA P2P approach has the best performance; it shows a nearly linear increase, and hence it is very scalable. This also demonstrates that DIANA is a suitable approach for large scale Grids and it can support increasing numbers of Grid nodes.

## 9. Related Work and Conclusions

Much work has been carried out in the domain of Grid scheduling but research in bulk scheduling and P2P scheduling for the Grid domain is relatively sparse. Prem et al. [9] present a P2P framework for the Grid but there is little work available on the queue management mechanism and their framework does not cover data intensive bulk scheduling. The CoreGRID Project [10] has worked for fault tolerant scheduling but they tackle mostly the compute operations whereas in DIANA we aim for data intensive scheduling. The DIANA P2P architecture is very much closer to the Napster architecture but Napster [11] is a P2P file sharing system whereas DIANA provides Grid enabled data intensive job scheduling. In the adaptive scheduling scheme [12] for data intensive applications, Shi et al, consider bandwidth as the only parameter for calculating data transfer cost. Moreover, they consider a deadline based scheduling approach, and the bulk scheduling issue is not covered. We have shown that other additional parameters not only need consideration in data intensive scheduling but that queues can be optimized by including these parameters in the decision criteria. The European Data Grid (EDG) Project has created a resource broker which is an extended and derived version of Condor but this is subject to the same issues and problems as Condor [13] itself. Although the problem of bulk scheduling has begun to be addressed in the most recent version of gLite the approach taken does not address network aware scheduling. Elmroth and Peterg [15] describe a Grid wide fair share scheduling system for local and global policies. They feature quota based scheduling and multilevel queues, although they do not consider reprioritisation, and it was not P2P oriented. The GridWay Scheduler [14] provides dynamic scheduling and opportunistic migration but its information collection and propagation mechanism is not robust, and it has not as yet been exposed to bulk job scheduling. The Gang scheduling [16] approach provides bulk scheduling by allocating similar tasks to a single location but it is tailored towards parallel applications working in a cluster whereas we are considering the meta-scheduling of the data intensive jobs submitted in bulk.

Our results indicate that considerable optimization can be achieved by applying P2P approaches to bulk scheduling. We have demonstrated that a scheduling cost-based approach can significantly improve the scheduling process if each job is submitted and executed after taking into consideration associated costs. Further details can be found in [6]. Our results demonstrate that a P2P meta-scheduler is better suited to Data Intensive and Network Aware (DIANA) scheduling than a single, centralized meta-scheduler. This paper demonstrated that if queue, priority and job migration are included in the bulk scheduling algorithm, the same algorithm could be used for the scheduling of bulk jobs. As a result, a multi-queue, priority-driven feedback based bulk scheduling algorithm is proposed and the results suggest that it can significantly improve the Grid scheduling and execution process. This not only reduces the overall execution and queue times of the jobs but also helps avoid resource starvation as well as creating a next generation scheduling platform for self organizing and decentralized scheduling of data intensive bulk jobs.